\documentclass[10pt,onecolumn]{article}
\usepackage{joss}
\usepackage{chemformula}

\usepackage[style=apa, backend=biber]{biblatex}
\jossRepoURL{https://github.com/dogusariturk/akaitools}
\jossArchiveDOI{10.5281/zenodo.19874850}
\jossDocsURL{https://dogusariturk.github.io/akaitools/}
\jossDate{Jun 2026}
\jossLogo{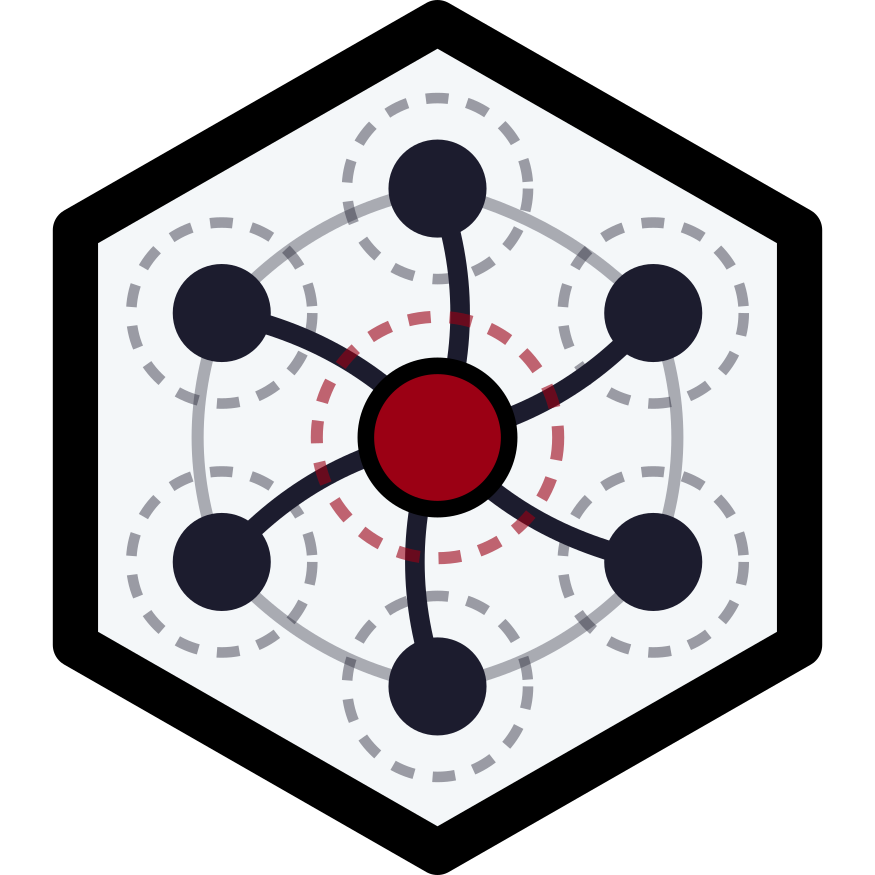}
\jossFooter{Sarıtürk \& Arróyave (2026). \textit{akaitools}: A Python package for parsing and analyzing AkaiKKR electronic structure calculations}

\jossauthor[1]{Doğuhan Sarıtürk}
\jossorcid{0000-0003-0008-1165}
\corrauthor{sariturk@tamu.edu}

\jossauthor[1,2,3]{Raymundo Arróyave}
\jossorcid{0000-0001-7548-8686}

\affiliation[1]{Department of Materials Science and Engineering, Texas A\&M University, USA}
\affiliation[2]{J. Mike Walker '66 Department of Mechanical Engineering, Texas A\&M University, USA}
\affiliation[3]{Wm Michael Barnes '64 Department of Industrial and Systems Engineering, Texas A\&M University, USA}

\title{akaitools: A Python package for parsing and analyzing AkaiKKR electronic structure calculations}

\begin{document}

\sloppy

\section{Summary}

The Korringa-Kohn-Rostoker (KKR) Green's function method \parencite{korringa1947,kohn1954} is a first-principles electronic structure approach well suited to substitutionally disordered alloys through the Coherent Potential Approximation (CPA) \parencite{soven1967}. AkaiKKR \parencite{akai1989} is a widely used implementation, known for efficient treatment of metallic systems and their magnetic properties. Its output, however, is unstructured plain text with no programmatic interface, leaving data extraction entirely to the user and making systematic or high-throughput studies impractical.

akaitools is a Python package that parses AkaiKKR output files into structured, type-annotated Python objects. The package covers three output types: self-consistent field (SCF) results, which capture convergence history and per-atom electronic and magnetic properties; spin-resolved, orbital-projected density of states for each CPA component; and Bloch spectral functions on a user-defined k-point path. Results come back as immutable dataclasses backed by NumPy arrays. Energy quantities are available in both Rydbergs and electronvolts, and results can be exported to Pandas DataFrames. A built-in plotting module produces Matplotlib figures for DOS curves and SCF convergence. A command-line interface provides file summaries and JSON export without any Python scripting. The package also includes a programmatic input file generator, so full calculation pipelines from input preparation to output analysis can be run in Python.

\section{Statement of Need}

AkaiKKR has been in use for decades, but structured programmatic access to its output remains largely manual. The official AkaiKKRPythonUtil \parencite{akaiutil} package parses output through individual methods, but they return a mix of scalars, lists, tuples, dicts, and DataFrames with no unified data model across output types. The higher-level EXPlotter layer couples parsing and figure generation in the same methods, so the parsed data cannot be used without also triggering rendering. The package is not on PyPI and is incompatible with current Python environments. The package ships no documentation site or tutorials. No existing materials science Python library covers AkaiKKR output.

akaitools is a Python 3.10+ package built around typed dataclass models for every parsed output quantity. When parsing fails, it raises an explicit exception identifying the file and the field that could not be read. DOS components can be filtered by element, site type, or spin channel, and the package detects associated spectral data files automatically. A command-line interface handles quick inspection and JSON export without any Python scripting. The test suite has over 60 cases run against real AkaiKKR output for Fe, Ni, NiFe alloys, GaAs, and Li, covering all three output types across spin-polarized and non-spin-polarized calculations.

\section{State of the Field}

The most direct precedent is AkaiKKRPythonUtil \parencite{akaiutil}, the Python utility package distributed by the AkaiKKR team. It parses self-consistent-field, DOS, and spectral function output through individual methods that extract total energy, RMS error per iteration, magnetic moments, atomic charges, Fermi level, and core-level energies. The methods return inconsistent types: scalars or plain lists from some, tuples, dicts, or Pandas DataFrames from others, with no shared data model across the three parsers. In the higher-level EXPlotter classes, parsing and figure generation are coupled in the same methods. Input files can be generated by passing a Python dict to a writer function, but the dict is not validated before writing, a debug print statement remains live in the code, and a known bug always writes k-path lines for spectral function inputs regardless of the dict contents. There is no documentation site, no API reference, and no tutorials. Installing it requires cloning the repository and running pip install from a subdirectory, as no PyPI release exists. Parts of the codebase do not work with current Python environments, and the package has no automated test suite or CI pipeline. No code has been pushed since September 2023; several open issues have gone unanswered. akaitools takes a different approach: all three parsers return from a single hierarchy of frozen, type-annotated dataclasses with shared base fields; parsing and visualization are independent layers; input generation uses a typed class that validates its contents and raises explicit errors; and the package ships a documentation site with usage guides, an API reference, and troubleshooting notes.

SPR-KKR \parencite{ebert2011}, developed by Ebert and colleagues at the University of Munich, is a fully relativistic KKR package with its own post-processing layer for transport properties, spectral functions, and optical conductivity. It is actively maintained and has a broad user base in relativistic electronic structure calculations. However, it is a separate code with a different input and output format from AkaiKKR. Its post-processing tools target SPR-KKR output specifically and cannot be applied to AkaiKKR files.

AiiDA-KKR \parencite{ruessmann2021} wraps the JuKKR code, a KKR implementation developed at Forschungszentrum Jülich, within the AiiDA workflow framework \parencite{huber2020}. It provides automated workflow management, provenance tracking, and high-throughput execution for Jülich's KKR implementation. Like SPR-KKR, it is a well-supported project for a different KKR code. The output format, binary file conventions, and CPA bookkeeping differ between KKR implementations, and AiiDA-KKR neither parses nor interfaces with AkaiKKR output.

\section{Software Design}

akaitools stores parsed results as frozen dataclass instances, not dictionaries or mutable objects. Values are fixed once a calculation output is read from disk. Workflows often pass the same result object through several analysis steps, and mutability in that position creates silent bugs. Named fields make the schema readable without opening the documentation; type annotations support IDE completion and static analysis. Frozen instances are also hashable, so they can serve as dictionary keys in lookup tables built across many calculations.

\begin{figure}
  \centering
  \includegraphics[width=\linewidth]{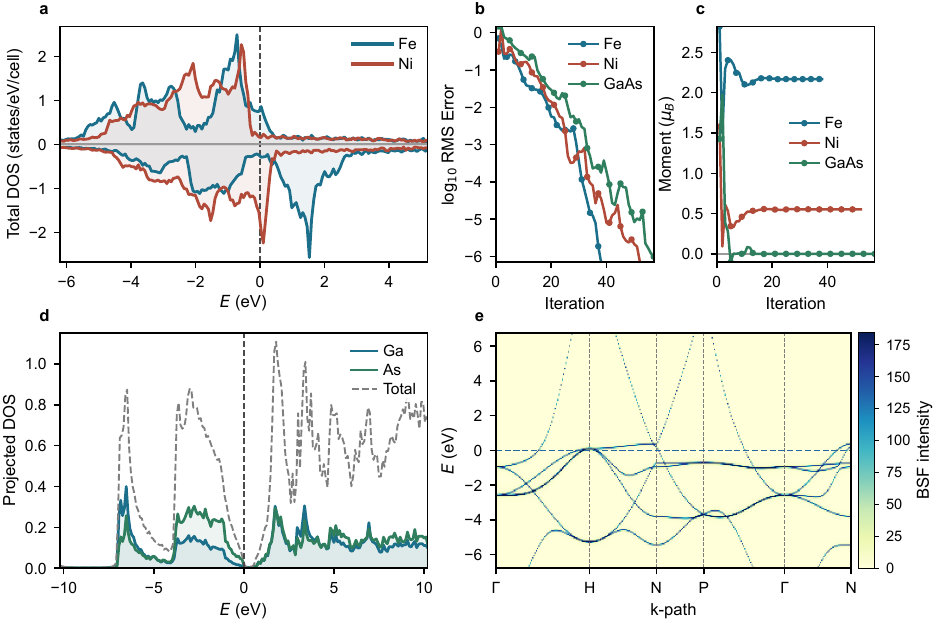}
  \caption{Electronic-structure results for representative metallic and semiconducting systems. In \textbf{a}, the spin-resolved total density of states of Fe and Ni is shown relative to the Fermi level, indicated by the vertical dashed line at $E = 0$~eV. Panel \textbf{b} reports self-consistent-field convergence histories for Fe, Ni, and GaAs in terms of the ($\log_{10}$) RMS error, and \textbf{c} shows the corresponding evolution of the magnetic moment during the iterative procedure. The GaAs electronic structure is examined further in \textbf{d}, where the projected density of states is decomposed into Ga and As contributions with the total DOS overlaid, and in \textbf{e}, which presents the band structure along the indicated high-symmetry k-path colored by Bloch spectral function intensity.\label{fig:figure1}}
\end{figure}

The package has four independent layers. Parsers extract text from raw AkaiKKR output; the model layer defines the typed schema; a utilities module holds shared Rydberg-to-eV constants used by both model properties and the plot module, with neither depending on the other; and the plot module takes model objects and returns Matplotlib \parencite{hunter2007} figures. \autoref{fig:figure1} shows representative output for Fe, Ni, and GaAs. The CLI wraps the parsers and writes JSON. Because the layers have no internal dependencies, adding a different visualization back end does not touch the parsers, and adding a new output type does not touch the plot module.

DOS and Bloch spectral function data are stored as NumPy \parencite{harris2020} arrays, not Python lists. This supports vectorized energy-window queries and zero-copy Matplotlib plotting; the arrays also pass directly to other scientific Python tools without conversion. The DOS component model is one CPA alloy component at one crystal site for one spin channel, not just one element. In a disordered binary alloy, each site carries contributions from each component; storing them in per-element arrays would collapse the CPA decomposition and discard information the method depends on. Users can query DOS components by element symbol, site index, or spin channel; \autoref{fig:figure1}\textbf{a} and \autoref{fig:figure1}\textbf{d} show total and projected DOS for Fe, Ni, and GaAs. DOS results export as a table of orbital-resolved curves indexed by component, spin, and energy; SCF iteration histories export with columns for charge neutrality, magnetic moment, total energy, and RMS error, as shown in \autoref{fig:figure1}\textbf{b} and \autoref{fig:figure1}\textbf{c}. \autoref{fig:figure1}\textbf{e} shows the Bloch spectral function for GaAs along a high-symmetry k-path; color encodes band character at each k-point and energy.

All three AkaiKKR output types share crystal structure, system metadata, and input parameters. A shared module extracts these fields once; each parser then calls that common extractor and extends the result with output-type-specific fields. This prevents the divergence seen in AkaiKKRPythonUtil, where the same field, such as the lattice parameter, was parsed with slightly different logic in different utilities. When parsing fails, the package raises an explicit exception naming the file, the output type, and the field that could not be read; it does not return null values or partial data.

The input file generator produces AkaiKKR input files from Python. Each run's output can feed directly into the input for the next without manual editing. Band-structure calculations take k-point path objects that encode high-symmetry sequences in fractional coordinates. Composition screening is a natural use case: a concentration sweep requires a valid, self-consistent input file at each alloy ratio, and the generator produces each one automatically.

Full type annotations throughout make the package compatible with static analysis tools. Continuous integration runs linting and the full test suite on every commit. The package ships on PyPI with a documentation site covering usage guides, an API reference, and troubleshooting notes. The input file generator and JSON CLI are built for integration with job schedulers and materials databases, so AkaiKKR calculations can slot into reproducible pipelines alongside other codes.

\section{Research Impact Statement}

The test suite runs against unmodified AkaiKKR output for elemental \ch{Fe} (BCC ferromagnet), \ch{Ni} (FCC ferromagnet), the disordered \ch{Ni_xFe_{1-x}} alloy system under CPA, \ch{GaAs}, and \ch{Li}. These span ferromagnetic metals, disordered alloys, semiconductors, and simple metals. The 60-plus test cases cover all three output types (GO, DOS, SPC), both spin-polarized and non-spin-polarized, with and without CPA alloy components.

AkaiKKR is used in studies of metallic magnetism: exchange interactions in Heusler alloys, magnetic anisotropy in transition-metal compounds, and electronic structure in dilute magnetic semiconductors. As high-throughput approaches to alloy design take hold in materials science, automated pipelines need structured access to KKR-CPA output. VASP and Quantum ESPRESSO have had output parsers for over a decade; AkaiKKR has not, which has kept it out of multi-code workflows. akaitools is the structured parser AkaiKKR has lacked.

CPA alloy composition screening previously required ad-hoc shell scripts and manual data extraction at each grid point. With akaitools, the same workflow is a Python loop: each iteration parses the previous output, generates the next input, and moves to the next composition. JSON export makes results portable across environments: a parsed calculation can be read, queried, and passed to a materials database or machine-learning pipeline without AkaiKKR installed.

\section{Acknowledgements}

R.A.\ and D.S.\ acknowledge support from the Army Research Office (ARO) through Grant No.\ W911NF-22-2-0117. Portions of this research were conducted with the advanced computing resources provided by Texas A\&M High Performance Research Computing.

\printbibliography

\end{document}